\documentstyle[sprocl]{article}
\def\Journal#1#2#3#4{{#1} {\bf #2}, #4 (#3)}
\def\CMP{\em Comm.\ Math.\ Phys.}

\def\NPB{{\em Nucl. Phys.} B}
\def\PLB{{\em Phys. Lett.}  B}
\def\PRL{\em Phys. Rev. Lett.} 
 
\def\PRD{{\em Phys. Rev.} D}
\def\ZPC{{\em Z. Phys.} C}

\def\ANP{\em Ann.\ Phys.\ }
\def\ijmpa{{\em Int.\ J.\ Mod.\ Phys.\ } A}

\def\be{\begin{equation}}
\def\ee{\end{equation}}
\def\bea{\begin{eqnarray}}
\def\eea{\end{eqnarray}}
\def\Tr{{\rm Tr }}
\def\lf{16\pi^2}
\def\llf{(16\pi^2)^2}
\def\lllf{(16\pi^2)^3}

\def\DRED{\ifmmode{{\rm DRED}} \else{{DRED}} \fi}
\def\DREDD{\ifmmode{{\rm DRED}'} \else{${\rm DRED}'$} \fi}
\def\NSVZ{\ifmmode{{\rm NSVZ}} \else{{NSVZ}} \fi}

\def \qq{\qquad}  
\def\pa{\partial}
\def\ga{\gamma}  
\def\de{\delta}  
\def\ep{\epsilon}
\def \la{\lambda}

\def\jbar{{\overline{j}}}

\def\Dbar{{\overline{D}}}

\def\alphadot{\dot\alpha}

\def\pslash{p\!\!\! /}
\def\gahat{\hat{\gamma}} 

\def\ephat{\hat{\epsilon}}
\def\ghat{\hat{g}}   

\def\Khat{\hat{K}}
\def\sy{supersymmetry}
\def\sic{supersymmetric} 
\def\ssm{supersymmetric standard model}
\def\psib{\overline{\psi}}

\def\epb{\overline{\epsilon}}
\def\lab{\overline{\lambda}}

\def\semi{;\hfil\break}
\def\mtilde{{\tilde m}}

\begin{document}
\rightline{}
\rightline{LTH 400}
\rightline{hep-ph/9707278}
\vskip1cm

\title{REGULARISATION OF SUPERSYMMETRIC THEORIES
\footnote{To appear in the book `Perspectives on
Supersymmetry', World Scientific, Editor G. Kane.}}
\author{ I.~JACK AND D.R.T.~JONES}
\address{Theoretical Physics Division, 
Chadwick Building, Peach Street, Liverpool,\\  L69 3BX, UK}
\maketitle\abstracts{
We discuss issues that arise in the regularisation 
of \sic\ theories.}

\section{Beyond the tree approximation}

In this chapter we consider issues of both practice and principle  that
arise when we take  \sic\ theories and calculate radiative  corrections.
It is usually the case that a symmetry of the  Lagrangian is still a
symmetry of the full quantum effective action; in which case we say that
radiative corrections preserve the symmetry.  There are important
exceptions to this rule, however: for example conformal invariance is in
general violated by radiative corrections,  and massless quantum
electrodynamics has a global $U_1$ axial symmetry  which is violated at
the one-loop level (in accordance with the famous Adler-Bardeen
theorem). It is not a priori obvious, therefore, that \sy\ is a symmetry
of the  full quantum theory in any particular case. Indeed it has been
occasionally claimed that there exists a \sy\ anomaly. In some cases
these  claims have been erroneous, and have occurred because it is
difficult to  distinguish between a genuine anomaly and an apparent
violation of a \sic\ Ward identity due  to use of a regularisation
method that itself violates \sy. Contrariwise, a detailed formal 
renormalisation program has been pursued in a series of papers by 
Piguet and collaborators~\cite{piguet}\ including one~\cite{pigueta}\ 
where a proof  that
\sy\ is {\it not\/} anomalous was presented. There is no real  
reason to doubt this conclusion (although the treatment of 
infra-red singularities in the program is a possible weakness), for 
the class of theories considered; the
evidence  adduced so far points to \sy\ being a symmetry of the full
quantum theory. (Note, however, recent suggestions that 
there may indeed exist a \sy\ anomaly 
in composite operators~\cite{dixon}.)

The existence of an anomaly is intimately related to the question of 
regularisation. Beyond the tree level, certain amplitudes in any  given
quantum field theory are not defined, due to divergences caused  by the
need to integrate over all momenta for particles in intermediate states.
 Regularisation is the process whereby the result of an ill-defined 
correction to a given amplitude is separated into a finite part (which
is  retained) and an ``infinite'' part (or more precisely, a part which
tends to  infinity in the limit that a certain parameter, or parameters,
specific  to the regularisation method is removed) which is removed from
the theory  (``subtracted'') by introducing a counter-term which 
precisely cancels it. If the regularised theory fails to  respect any
given symmetry then the finite amplitudes will fail to satisfy  the Ward
identities of the symmetry, giving rise to an apparent anomaly  and the
confusion alluded to above. When the anomaly really is specious,  it is
possible to restore invariance by modifying the counter-terms  by finite
amounts. (Obviously the counter-terms are ambiguous, in that  their
defining role is to cancel something which becomes infinite as  the
regulator is removed, so that adding a finite quantity  to any
counter--term leaves its {\it raison d'\^etre\/} intact.  If no
modification of the counter-terms  will restore the Ward identity then
there is an anomaly.\footnote{Sometimes an anomaly can be apparently
removed only to  reappear in another guise. This is the case with the
Adler--Bardeen  anomaly, which is a property of the fermion triangle
with two vector  and one axial--vector vertices. The anomaly may affect
the axial current  or the vector current, depending on how the theory is
regularised.}

From a formal point of view, the choice of 
regularisation scheme made in the implementation of a renormalisation 
is not of great significance; it is important only that it 
corresponds to addition of {\it local\/} counter-terms. 
For the extraction of physical predictions, however, the 
choice becomes a matter of considerable practical significance. 
It is convenient, for example, to use a regularisation
method that preserves  symmetries. It should be clear, in fact, from the
above discussion that the existence of a regulator consistent with 
a given symmetry suffices to prove that symmetry to be anomaly--free. 
In this context the approach of West~\cite{westb}, 
consisting of higher derivative regularisation 
supplemented by Pauli-Villars~\footnote
{Use of Pauli-Villars at one loop in supersymmetry was also 
advocated by Gaillard~\cite{maryk} in 1995.}   
at one loop,  is worthy of consideration; 
but as the author himself remarks, the issue of a possible anomaly 
is not thereby fully resolved because of the infra-red difficulties already 
alluded to.

Dimensional regularisation (DREG) is an elegant and convenient way of 
dealing with the infinities that arise in  quantum field theory beyond
the tree approximation.~\cite{hvelt}  It is well adapted to gauge
theories because it preserves gauge invariance;  it is less well-suited,
however, for supersymmetry  because invariance of an action  with
respect to supersymmetric transformations only holds in general for 
specific values of the space-time dimension $d$.   
This is essentially due to the fact that a
necessary  condition for supersymmetry is  equality of Bose and Fermi
degrees of freedom.   In non-gauge theories it  is relatively easy to
circumvent this problem,  and DREG as usually employed is, in fact, a
supersymmetric procedure.  Gauge theories are a  different matter,
however, and the question as to whether there  exists a completely
satisfactory supersymmetric  regulator for gauge theories remains 
controversial. This fact has been exploited recently to suggest that
there  may be  supersymmetric anomalies.~\cite{dixon} 

An elegant attempt to modify DREG so as to render it  compatible with
supersymmetry was  made by Siegel.\cite{siegel} The essential
difference between Siegel's  method (DRED) and DREG is that the
continuation from $4$ to $d$ dimensions  is made by {\it
compactification\/}, or {\it dimensional reduction}.  Thus  while the
momentum (or space-time) integrals are $d$-dimensional in the  usual
way, the number of field components remains unchanged and consequently 
supersymmetry is undisturbed. (A pedagogical introduction to \DRED
was given by Capper et al~\cite{Capper}.)

As pointed out by Siegel himself,~\cite{siegelb} there remain
potential  ambiguities with \DRED associated with treatment of the
Levi-Civita symbol, $\epsilon^{\mu\nu\rho\sigma}$. We will address 
this difficulty and the related one involving $\ga^5$ in 
Section~\ref{sec:prob}.

We must also address problems   which arise only when \DRED   is applied
to non-supersymmetric theories.  That \DRED is a viable alternative to
DREG in the non--\sic\ case   was claimed early on~\cite{Capper}. 
Subsequently it has been adopted occasionally,  motivated usually by the
fact that Dirac matrix algebra is easier  in four dimensions--and in
particular by the desire to use 
Fierz identities~\cite{Korner}$^{\!,\,}$\cite{Misiak}. 
One must, however,  be very
 careful in applying \DRED to non-supersymmetric  theories because of the
existence of  {\it evanescent couplings}. These were first 
described~\cite{Tim}\ in 1979, 
and  independently discovered by van Damme and 't Hooft~\cite{hvand}. 
They argued, in fact,  that while \DRED is a
satisfactory procedure for supersymmetric  theories 
\cite{jack}$^{\!,\,}$\cite{jacko} 
(modulo the  subtleties alluded to above) it leads to a
catastrophic loss of unitarity in  the non-supersymmetric case.
Evidently there is an important  issue to be resolved here--is 
use of \DRED  in fact forbidden (except in the
supersymmetric case) in spite  of its apparent convenience? It has been
conclusively  demonstrated~\cite{dreda}$^{\!,\,}$\cite{dredb} that if \DRED is
employed in the manner  envisaged by Capper et al~\cite{Capper}, (which
as we shall see differs in an important  way from the definition of \DRED
primarily used by `t Hooft and van Damme) then there is no problem  with
unitarity. There exist  a set of transformations  whereby the
$\beta$--functions of a particular theory (calculated using DRED) may  be
related to the $\beta$--functions of the same theory (calculated using DREG)
by  means of coupling constant reparametrisation. The key is that a
correct description  of any non-supersymmetric  theory impels us to a
recognition of the fundamental fact  that in general the evanescent
couplings renormalise in a manner different  from the
``real'' couplings with which we may be tempted to associate them. This
means  that care must be taken as we go beyond one loop; nevertheless it
is still possible  to exploit the simplifications in the Dirac algebra
which have motivated the use of DRED.  We will return to this point
later. 

At this point the reader may wonder why, in a book about \sy, we should
worry about  renormalising non--\sic\ theories at all. The main 
practical reason is
that the \ssm\ is an effective
theory in which \sy\ is {\it explicitly\/}  broken, albeit by terms with
non--zero dimension of mass.   

The reader may also feel that, given the problems with $\DRED$, 
we should explore other regulators. For example, there has been 
some recent work on a new approach known as differential 
regularisation~\cite{dan}. The fact is, however, that 
the convenience of DREG for calculations beyond one loop makes 
the use of some variant of it very desirable. Use of other 
proposed regulators is rarely pursued beyond verification 
of some already known (and usually one-loop) results. 

\section{Introduction to \DRED}\label{sec:intro}

As a concrete example, let us consider  a non-abelian gauge theory with
fermions but no elementary  scalars. The theory to be studied consists
of a Yang-Mills multiplet $W^a_{\mu} (x)$ with  a multiplet of spin
$1\over 2$ fields~\footnote{which may be Dirac or Majorana at this stage} 
$\psi^{\alpha}(x)$ transforming
according to  an irreducible representation $R$ of the gauge group $G$. 
Of course if $\psi$ {\it is\/} Majorana, then $R$ must be a real 
representation, since the Majorana condition is not preserved by a unitary 
transformation. 

The Lagrangian density (in terms of bare fields) is 
\begin{equation}
L_B = -{1\over 4}G^2_{\mu\nu} - {1\over{2\alpha}}(\pa^{\mu}W_{\mu})^2 + 
C^{a*}\pa^{\mu}D_{\mu}^{ab}C^b 
+ i\psib^{\alpha}\ga^{\mu}D_{\mu}^{\alpha\beta}\psi^{\beta} 
\label{eq:AA} 
\end{equation}
where 
\begin{equation}
G^a_{\mu\nu} = \pa_{\mu}W_{\nu}^a 
- \pa_{\nu}W_{\mu}^a + gf^{abc}W_{\mu}^b W_{\nu}^c
\end{equation}
and 
\begin{equation}
D_{\mu}^{\alpha\beta} = \de^{\alpha\beta}\pa_{\mu} 
- ig (R^a )^{\alpha\beta}W_{\mu}^a
\end{equation}
and the usual covariant gauge fixing and ghost terms have been introduced. 

The process of dimensional reduction consists of imposing that all field
variables depend only  on a subset of the total number of space-time
dimensions-- in this case $d$ out of $4$ where  $d = 4 - \ep$. 
We will use $\mu, \nu\cdots$ to denote $4$-dimensional indices and 
$i, j$ to denote $d$-dimensional ones, with corresponding 
metric tensors $g_{\mu\nu}$ and $g_{ij}$. It is also convenient 
to introduce ``hatted'' quantities (such as $\ghat_{\mu\nu}$ and 
$\gahat^{\mu}$) which are identical to the corresponding 
$d$-dimensional quantities ($g_{ij}, \ga^{i}\cdots$) within the 
$d$-dimensional subspace, but whose remaining components are zero. 
Momenta $p_{\mu}$ exist only in the 
$d$ dimensional subspace so we do not bother to ``hat'' them. 
Thus we have for example 
\be
\pslash = p_{\mu}\ga^{\mu} = p_{\mu}\gahat^{\mu}
\ee
and
\be
g^{\mu\nu}g_{\mu\nu} = 4, \quad 
\ghat^{\mu\nu}\ghat_{\mu\nu} = g^{ij}g_{ij} = d.
\ee
In particular, we have that
\be
\ghat^{\mu\nu}g_{\nu}{}^{\la} = \ghat^{\mu\la}\qq \hbox{and}\qq
\ghat^{\mu\nu}\ga_{\nu} = \gahat^{\mu}.
\label{eq:pba}
\ee
These apparently innocuous relations will cause us trouble 
in the next section.   

In order to fully appreciate the consequences of \DRED for $L_B$ we must
make the decomposition 
\begin{equation}
W_{\mu}^a(x^j ) = \{ W_i^a (x^j ), W_{\sigma}^a(x^j )\} \label{eq:AB}
\end{equation}
where
\begin{equation}
\de^i_i = \de^j_j = d \qq \hbox{and}\qq \de_{\sigma\sigma} = \ep.
\end{equation}
It is then easy to show that
\begin{equation}
L_B = L _B^d + L_B^{\ep} \label{eq:AC}
\end{equation}
where
\begin{equation}
L _B^d = -{1\over 4}G^2_{ij} -{1\over{2\alpha}}(\pa^{i}W_{i})^2 +
C^{a*}\pa^{i}D_{i}^{ab}C^b 
+ i\psib^{\alpha}\ga^{i}D_{i}^{\alpha\beta}\psi^{\beta}
\label{eq:AD}
\end{equation}
and 
\begin{equation}
 L_B^{\ep} = {1\over 2}(D_i^{ab}W^b_{\sigma})^2 
- g\psib\ga_{\sigma}R^a\psi W_{\sigma}^a
-{1\over 4}g^2 f^{abc}f^{ade}W^b_{\sigma}W^c_{\sigma'}W^d_{\sigma}
W^e_{\sigma'}.
\label{eq:AE}
\end{equation}

Conventional dimensional regularisation (DREG) amounts to using 
Eq.~\ref{eq:AD} and discarding  Eq.~\ref{eq:AE}. For \DRED, on the other 
hand we include both.~\footnote
{The additional contributions from $L_B^{\ep}$ are precisely 
what is  required to restore the supersymmetric
Ward identities at  one loop in supersymmetric theories, as  described 
in section~\ref{sec:ward}.}
In simple  applications it is in general
more  convenient to eschew the separation performed above  and 
calculate with 4-dimensional  and $d$-dimensional indices rather than
$d$-dimensional and $\ep$-dimensional ones. As a simple illustration, 
consider the following typical calculation:
\be
\ga^{\mu}\pslash\ga_{\mu} = p_{\nu}\ga^{\mu}\ga^{\nu}\ga_{\mu} = 
-2p_{\nu}\ga^{\nu} = -2\pslash
\ee
or equivalently
\be
\ga^{\mu}\pslash\ga_{\mu} = \ga^{i}\pslash\ga_{i} + 
\ga^{\sigma}\pslash\ga_{\sigma} = (2-d)\pslash + (d-4)\pslash = -2\pslash.
\ee

From the dimensionally reduced form of the
gauge transformations:
\begin{equation}\begin{array}{rcl}
\de W^a_i & = &\pa_i\Lambda^a + gf^{abc}W^b_i\Lambda^c \\ 
\de W^a_{\sigma} & = & gf^{abc}W^b_{\sigma}\Lambda^c \\
\de\psi^{\alpha} & = & ig(R^a )^{\alpha\beta}\psi^{\beta}\Lambda^a \\
& & \end{array}
\label{eq:AF}
\end{equation}
we see that each term in Eq.~\ref{eq:AE} is separately invariant under gauge 
transformations. The $W_{\sigma}$-fields behave exactly like scalar 
fields, and are hence known as $\ep$-scalars.   The significance of this
is that gauge invariance {\it per se\/} provides 
no  reason to expect the $\psib\psi
W_{\sigma}$ vertex to renormalise in the same  way as the $\psib\psi
W_i$ vertex. In the 
case of the quartic $\ep$-scalar coupling the situation is more complex
since  in general of course more than one such coupling is permitted by
Eq.~\ref{eq:AF} . In  other words, we cannot in general expect the
$f-f$~tensor structure  present in Eq.~\ref{eq:AE} to be preserved under
renormalisation. This is clear from the abelian  case, where there is no
such quartic interaction in $L_B^{\ep}$ but there is  a divergent
contribution at one loop from a fermion loop.       

In the case of  \sic\ theories, however, these 
difficulties  do not arise. If $\psi$ above is a Majorana fermion 
in the adjoint representation, then $L_B$ is \sic. 
This links $W_i$ and $W_{\sigma}$ in a way that is not severed by the 
dimensional reduction. Thus the $\psib\psi
W_{\sigma}$ and $\psib\psi
W_i$ vertices (both equal to $g$ at the tree level) remain 
equal under renormalisation. 
We will return in section~\ref{sec:nonsic}\ to the application of \DRED 
to non-supersymmetric theories. 

\section{\DRED ambiguities}\label{sec:prob}

With \DRED it would seem that necessarily $d<4$,
since the regulated action is, after all, defined by  dimensional {\it
reduction}. 
Then, given $d<4$, one can define an  object
$\ephat^{\mu\nu\rho\sigma}$ as follows: 
\be
\ephat^{\mu\nu\rho\sigma} = 
\ghat^{\mu\alpha}\ghat^{\nu\beta}\ghat^{\rho\ga}\ghat^{\sigma\delta}
\epsilon_{\alpha\beta\ga\delta}
\label{eq:probb}
\ee
where $\epsilon_{\alpha\beta\ga\delta}$ is the usual $4$-dimensional 
tensor. Unfortunately it is now possible to show 
that algebraic inconsistencies result~\cite{siegelb} unless $d=4$.
Let us illustrate these problems in the two dimensional case. The 
alternating tensor $\epsilon^{\mu\nu}$ satisfies (in two Euclidean dimensions)
the relation
\be
\epsilon^{\mu\nu}\epsilon^{\rho\sigma} = 
g^{\mu\rho}g^{\nu\sigma} - 
g^{\mu\sigma}g^{\nu\rho}. 
\ee
Using Eq.~\ref{eq:pba}\ it is easy to show that 
\be
\ephat^{\mu\nu}\ephat^{\rho\sigma} = 
\ghat^{\mu\rho}\ghat^{\nu\sigma} - 
\ghat^{\mu\sigma}\ghat^{\nu\rho}
\label{eq:probc}
\ee
where $\ephat^{\mu\nu}$ is defined similarly to Eq.~\ref{eq:probb}.

However it is trivial to 
demonstrate that the result of 
applying Eq.~\ref{eq:probc} to the tensor 
\be
A^{\mu\nu}=\ephat^{\mu\nu}\ephat^{\rho\sigma}\ephat_{\rho\sigma}
\label{eq:epsa}
\ee 
is ambiguous inasmuch that  it differs according to which pair of 
$\ephat$-tensors are selected: the result is the identity
\be
(d + 1) ( d - 2) \ephat^{\mu\nu} = 0.
\ee 
 
A  related problem (of course) is the fact that the only 
mathematically consistent 
treatment of $\ga^5$ within DREG is predicated~\cite{hvelt}
on having $d>4$. Given Eq.~\ref{eq:pba} and the usual relation
\be 
\left\{ \ga_{\mu}, \ga^5\right\}=0.
\label{eq:fivea}
\ee
it follows that 
\be
\left\{ \gahat_{\mu}, \ga^5\right\}=0.
\label{eq:fiveb}
\ee
and hence that 
\be
(d-4)
\Tr\left[\ga^5\gahat^{\mu}\gahat^{\nu}\gahat^{\rho}\gahat^{\sigma}\right]=0. 
\ee
This is unfortunate  since it renders problematic the discussion 
of the axial anomaly.

For $d>4$, however, Eq.~\ref{eq:pba} does not hold and so 
Eq.~\ref{eq:fiveb} no longer follows. 
Instead we impose
\be
\left[ \ga_{\sigma}, \ga^5\right]=0, \quad\hbox{for}\quad 4<\sigma<d
\label{eq:fivec}.
\ee
and this leads to a 
straightforward and unambiguous derivation of the axial anomaly. 
It has been verified~\cite{Leveille} 
that this prescription correctly reproduces the 
Adler-Bardeen theorem at the two-loop level using DREG.

Returning to the \DRED prescription, there are a number of possible 
``fixes'' at one loop;~\cite{nict} at two loops, 
it was shown~\cite{Leveille} that 
the Adler-Bardeen theorem could indeed still be satisfied if 
relations like
\be
\ga^{i}\ga^5\ga{i} = (d-8)\ga^5
\ee
which follow in the $d>4$ case, are 
used in conjunction with \DRED. 

A possible point of view concerning all this~\cite{bonneau} is that \DRED 
is terminally inconsistent and should not be used. We believe, 
however, that the difficulties are essentially technical and can be 
evaded. For example, one well-defined procedure would be to 
write 
\be
\ga^5 = \frac{1}{4!}\ep^{\mu\nu\rho\sigma}
\ga_{\mu}\ga_{\nu}\ga_{\rho}\ga_{\sigma}
\ee
and factor all out $\ep$-tensors. Renormalised amplitudes may then be 
calculated, which, being finite as $d\to 4$, are unambiguous when 
the $\ep$-tensors are contracted in. 
\footnote{This somewhat cumbersome procedure
can usually be finessed. For instance, if only even-parity 
fermion loops are present, then there is no problem with a 
fully anti-commuting $\ga^5$.}
We would claim also that other 
modes of procedure which would give different answers because of the 
ambiguities detailed above, correspond nevertheless to {\it the same 
physical results}. This assertion has in fact been verified in 
one particular case~\cite{allen}, where a prescription first suggested 
by Hull and Townsend~\cite{hullta} was used. This amounted to employing as  
$\ep^{\mu\nu}$ not the usual alternating tensor but instead a 
structure satisfying 
\be
\ephat^{\mu}{}_{\nu}\ephat^{\nu\rho} = (1 + c\epsilon)\ghat^{\mu\rho}
\ee
(where here $\epsilon = 2-d$). It turns out~\cite{allen} 
that the dependence of 
the results on the parameter $c$ can be absorbed into redefinitions 
of the renormalised metric and torsion tensors. In the special case 
$c=0$, $\ephat^{\mu\nu}$ is an 
almost complex structure~\cite{osborn}. 

There have been a considerable  number of papers discussing the 
interpretation of $\ga^5$ in both DREG and \DRED, and the reader may 
consult them for further enlightenment~\cite{gfives}. 
We turn in the next section to   another (but again related) 
problem with \DRED, arising from 
the fact that in spite of the correct counting of degrees of freedom, 
there are still ambiguities associated with  establishing invariance of
the action: we will look at this in more
detail below in the context  of the \sy\ Ward identity.
 
\section{The supersymmetry Ward identity}\label{sec:ward}

The first concrete illustration of the different results provided 
by \DRED and DREG for a supersymmetric theory was as follows.  
Consider the basic \sic\ gauge theory in the Wess-Zumino gauge as defined 
by the Lagrangian $L_S$ where
\be
L_S = -{1\over 4}G^{\mu\nu}G_{\mu\nu} + 
i\frac{1}{2}\lab^{\alpha}\ga^{\mu}D_{\mu}^{\alpha\beta}\la^{\beta} 
+ {1\over 2}D^2.
\label{eq:EB}
\ee
In $d = 4$, $L_S$ is invariant (up to a total derivative) under the 
transformations
\bea
\de W^a_{\mu} & = & i\epb\ga_{\mu}\la^a, \qq \de\la^a = 
{1\over 2}G^a_{\mu\nu}\ga^{\mu}\ga^{\nu}\ep -iD^a\ga^{5}\ep
\nonumber \\
\de D^a & = & -\epb\ga_{\mu}\ga^{5}(D^{\mu}\la )^a.\label{eq:EBA}  
\eea
It is an excellent exercise in spinor algebra to verify this invariance. 
Note the presence in Eq.~\ref{eq:EBA}\ of 
$\ga^{5}$-terms; to obtain invariance 
one must assume that $\ga^{5}$ is totally anti-commuting. Of course in this 
particular case we could set $D^a=0$, and still have an invariance (not 
involving $\ga^{5}$). However 
this does not escape the Siegel ambiguity, as we shall now show.   
With due care, one obtains (up to total derivatives) 
\be
\delta L_S = 
g\frac{1}{2}f^{abc}\epb\ga^{\mu}\la^a\la^b\ga_{\mu}\la^c. \label{eq:EBB}
\ee
This is identically zero for $d=4$, though this is not obvious  
even if we rewrite in two--component formalism; a Fierz re-ordering is 
required.  For $d\neq 4$, $\delta L_S$ 
is not zero; and the key to the distinction between \DRED and DREG  
lies in the $\ga^{\mu}\otimes\ga_{\mu}$ contraction, which is 
$d$-dimensional for DREG and four-dimensional for \DRED. 
There are  important consequences for the regularisation 
of \sic\ theories, as we shall now see.  
Let us add to $L_S$ gauge fixing and ghost terms: 
\begin{equation}
L_S\to L_T = L_S + L_G
\ee
where
\be
L_G =  - {1\over{2\alpha}}(\pa^{\mu}W_{\mu})^2 + 
C^{a*}\pa^{\mu}D_{\mu}^{ab}C^b.\label{eq:EBC}
\end{equation}
Then we introduce the functional $Z(J, j, j_D )$ where 
\begin{equation}
Z = \int d\{W_{\mu}\} d\{\la\}d\{ D\} e^{i\int d^d x\left[ 
L_T + J^{\mu}W_{\mu} + \jbar\la + j_D D\right]}\label{eq:EBD}
\end{equation}
and use of Eq.~\ref{eq:EBA} leads to the following 
Ward identity: 
\begin{equation}
0 = <\int d^d x \left[ J^{\mu}\de W_{\mu} + \jbar\de\la + j_D \de D 
+ \de L_S + \de L_G\right]>\label{eq:EBE}
\end{equation}
where 
\be
\de L_G = - {1\over{\alpha}}\pa^{\mu}W_{\mu}\pa^{\mu}\de W_{\mu} 
+ gf^{abc}C^{a*}\pa^{\mu}\de W_{\mu}^{c}C^b\label{eq:EBF}
\ee
and  
\begin{equation}
<X> = \int d\{W_{\mu}\} d\{\la\}d\{ D\} X e^{i\int d^d x\left[ 
L_T + J^{\mu}W_{\mu} + \jbar\la + j_D D\right]}.\label{eq:EBG}
\end{equation}
Notice we have included the term $\de L_S$ from Eq.~\ref{eq:EBB} 
to allow for the fact that this may not be zero away from $d=4$.
When this Ward identity was investigated at one loop~\cite{Capper},  it
was found to be true with \DRED and false with DREG; which conclusion was
 arrived at because the contribution from $\de L_S$ was ignored. It  is
easy to show that this contribution is zero for \DRED, and  in the case
of DREG serves precisely to restore Eq.~\ref{eq:EBF}.
The distinction between \DRED 
and DREG is manifest in the fact that the contribution of 
$\de L_S$ is zero in the former case. It is in this sense that \DRED 
is more consistent with \sy. In terms of this Ward identity the difference 
may not seem crucial, but the fact that $\de L_S$ is effectively non-zero 
(with DREG) means that if we employ DREG in a \sic\ theory then great care 
must be taken with the formulation of physical predictions. 
This becomes particularly clear when we generalise to include matter fields; 
in \sic\ QCD, for example, the fact that the 
gluino-quark-squark coupling is equal to the
 gauge coupling $g$ is a consequence of \sy\ and will not be preserved 
under renormalisation if DREG is employed. In fact, this is very similar 
to the problems that occur when we want to apply \DRED 
to non--\sic\ theories; once again there are coupling constant relations 
that are not preserved by radiative corrections.    

While \DRED was successful in the above application, it does not follow that 
an insertion of $\de L_S$ in a diagram of arbitrary complexity gives 
zero. It can be shown~\cite{vlad}
that such an insertion depends on the quantity 
$\Delta$ where 
\begin{equation}
\Delta = \Tr (A \ga^{\mu} B \ga_{\mu})  + \Tr (A \ga^{\mu}) \Tr (B \ga_{\mu} ) 
- (-1)^k \Tr (A \ga^{\mu} B^R \ga_{\mu}) .
\end{equation}
Here $A$ and $B$ are products of Dirac $\ga$-matrices; $k$ is the number of 
such matrices in $B$, and $B^R$ consists of the the same set of 
matrices as $B$ but written down in reverse order. In strict $d=4$, 
$\Delta$ is zero; but because $A$, $B$ may contain $d$-dimensional 
$\ga$-matrices (due to contraction with momenta) $\Delta$ is 
non-zero in general. If we set 
\begin{equation}
A = \ga^{\mu_1}\ga^{\mu_2}\cdots \ga^{\mu_5}   
\quad\hbox{and}\quad B= \ga_{\nu_1}\ga_{\nu_2}\cdots \ga_{\nu_5} 
\end{equation}
then 
\begin{equation}
\Delta = 48 \de^{[\mu_1}_{\nu_1}\de^{\mu_2}_{\nu_2}
\cdots\de^{\mu_5 ]}_{\nu_5}
\end{equation}
where the brackets $[\cdots ]$ denote antisymmetrisation. 
This is clearly zero for integer  $d\leq 4$ but if the various indices are
$d$-dimensional  then it is not. For instance if we calculate the trace
by contracting  $\Delta$ with $\de_{\mu_1}^{\nu_1}\cdots \de_{\mu_5}^{\nu_5}$
then  we obtain $\Tr\Delta = 48 d (d-1) (d-2) (d -3) ( d-4)$. 
A diagram with at least
four loops is required~\cite{vlad} so that we get enough $\ga$-matrices 
to activate this problem in the propagator Ward identity~\cite{Capper}. 
This is clearly the same ambiguity at bottom  
addressed by Siegel~\cite{siegelb}. 

One might hope that this problem is somehow resolved by  use of
superfields. Using superfield perturbation theory  Feynman rules
maintains \sy\ in a manifest way; but a crucial part of the 
calculational procedure relies on the reduction of products of
supercovariant  $D_{\alpha}$ and $\Dbar_{\alphadot}$ derivatives to
products of four or less,  and this is possible only when the fact that
the $\alpha, \alphadot$ indices  are two-valued is used. Thus the same
ambiguity must be present,  albeit in a somewhat different form. 
However as we argued in the previous section, the ambiguity  will not
affect physical results since it is equivalent to the freedom available
in choice of regularisation scheme, as  long as a systematic procedure
is adopted. Despite all difficulties, \DRED remains the regulator of
choice for supersymmetric theories, and  has survived many practical
tests. 

\section{$N=2$ and $N=4$ \sy}\label{sec:moren}

$N=2$ \sy\ corresponds, in the language of $N=1$ superfields, to the 
special case of a superpotential taking the form: 
\be
W=\sqrt{2}g\phi^a\xi^TS_a\chi,
\ee
where $\xi, \chi, \phi$ are multiplets transforming under the
$S^*$, $S$ and adjoint representations of the gauge group ${\cal G}$ 
respectively. In the special case that $S$ is the adjoint representation 
we have $N=4$ \sy. $N=2$ theories are extraordinary in that they 
have only one-loop divergences~\cite{hsw}. This means that for $N=2$, 
$\beta_g$ vanishes 
beyond one-loop if computed using $\DRED$; crucial here is the fact 
that DRED incorporates {\it minimal subtraction}.  In the $N=4$ case the 
one-loop contribution also vanishes, so $N=4$ theories are 
ultra-violet finite to all orders of perturbation theory. 

$N=2$ and $N=4$ theories, although obviously of great 
interest, possess a property that is unfortunate if we want  to try and
incorporate them into a realistic theory. This property is  that since the
chiral superfields are either adjoint or  in $S, S^*$ pairs, gauge
invariant mass terms are possible for  the fermionic components of the
multiplets. It is difficult,  therefore, to arrange for fermion masses
(such as the electron  mass) to be much less than the scale of
\sy--breaking, at least. Nevertheless there have been occasional attempts 
to construct phenomenologically viable models~\cite{aggy}, and 
explore their consequences~\cite{anti}. 
 
\section{Non--\sic\ theories}\label{sec:nonsic}

We saw in section~\ref{sec:intro}\ that under renormalisation the 
$\epsilon$-scalars behave differently from the gauge fields, except  in
\sic\ theories. On might be tempted to assert that 
it doesn't matter if  Green's
functions with external $\ep$-scalars are divergent (since they are
anyway unphysical) and introduce a common wave function subtraction for
$W_i$ and $W_{\sigma}$, a wave function  subtraction for $\psi$ and a
coupling constant subtraction for $g$, these being determined  (as
usual) by the requirement that Green's functions with real particles be
rendered finite.  This was the procedure adopted in the main by 
van~Damme and 't~Hooft~\cite{hvand}.  On the other hand  we could insist
on all Green's functions (including those with external $\ep$-scalars) 
being finite,  leading to the introduction  of a plethora of new
subtractions or equivalently coupling constants.  We have 
shown~\cite{dreda}$^{\!,\,}$\cite{dredb} that
 it is only the latter procedure which leads to a
consistent theory; the  former manifestly breaks 
unitarity.

Now in a \sic\ theory the complications  described above can be safely
ignored. The wave function renormalisations of    $W_{\sigma}$ and
$W_i$ are equal because of \sy, and  the evanescent
couplings remain equal to their ``natural'' values after 
renormalisation. At first sight, this conclusion also  appears to obtain
when \sy\ is softly broken, since the dimensionless  couplings
renormalise exactly as in the fully \sic\ theory. This  is not quite
true, however, since there is nothing to protect  the $\ep$--scalars
acquiring a mass  through interacting with  the genuine fields; 
and, indeed, precisely this happens~\cite{jj}.  
In other words, the $\beta$--function for the $\ep$-scalar mass $\mtilde$ 
is inhomogeneous with respect to $\mtilde$:
\be
\beta_{{\mtilde^2}} = A(g, Y)\mtilde^2 + \sum_i B_i(g, Y)m_i^2 + \cdots, 
\ee
where the $m_i^2$ are the genuine  scalar masses, 
$Y$ represents the Yukawa couplings and the $+\cdots$ denotes terms  
involving  the gaugino mass(es) and the $A$-parameter(s).  Moreover, the
two--loop $\beta$--functions for the genuine scalar masses  
depend explicitly on
the $\ep$--scalar masses, when calculated using $\DRED$.  This fact would
annoyingly complicate an extension to two loops of  the 
standard running analysis relating the low energy values of the various soft 
parameters to the corresponding values at gauge unification. 
Fortunately, however, there
exists a hybrid scheme~\cite{jjmvy}  
which decouples this $\ep$--scalar dependence
both from the $\beta$--functions  and from the threshold corrections to
the physical masses. At leading order, this scheme is arrived 
at from \DRED by redefining the masses $m_i^2$ as follows:
\be
m_i^2 |_{{\rm DRED}'} = m_i^2 |_{{\rm DRED}} - C_i (g)\mtilde^2
\ee
where $C_i (g)$ is easily calculated~\cite{jjmvy}. The resulting 
$\beta_{m^2}$ is independent of $\mtilde^2$ through two loops; and 
conveniently the same transformation   removes the $\mtilde^2$ term 
from the one-loop relationship between the renormalised and physical  
scalar masses $m^2$. So in the ${\rm DRED}'$ scheme, although $\mtilde$ 
evolves under the renormalisation group, it is decoupled from 
physical quantities and so can be safely ignored. To sum up, the 
${\rm DRED}'$ procedure for the standard running analysis (at two loops) 
and extraction of predictions for physical masses is:

(1) Use the ${\rm DRED}'$ $\beta$-functions~\cite{jj}$^{\!-\,}$\cite{yam}
to do the running analysis.

(2) Calculate the one loop corrections to convert the 
renormalised masses to the physical (pole) masses using \DRED but with 
$\mtilde^2 = 0$. The top quark mass would not in any case have $\mtilde$ 
dependence, but note that the result for the one--loop 
gluon contribution to it is  
\be
m_t^{\rm pole} = m_t (\mu ) \left[1+\frac{\alpha_3 (\mu)}{3\pi}
\left(5 - 3\ln\frac{m_t^2}{\mu^2}\right)\right]
\ee  
not 
\be
m_t^{\rm pole} = m_t (\mu ) \left[1+\frac{\alpha_3 (\mu)}{3\pi}
\left(4 - 3\ln\frac{m_t^2}{\mu^2}\right)\right]
\ee  
as in DREG.

\section{The \NSVZ $\beta$-function}\label{sec:nsvz}

In this section we examine the $\beta$-functions 
for an $N=1$ \sic\ theory
defined by the superpotential
\be
W=\frac{1}{6}Y^{ijk}\Phi_i\Phi_j\Phi_k+\frac{1}{2}\mu^{ij}\Phi_i\Phi_j.
\ee
The multiplet of chiral superfields $\Phi_i$ transforms as a
representation $R$ of the gauge group $\cal G$, which has 
structure constants $f_{abc}$. 
In accordance with the non-renormalisation theorem~\cite{gsr}, 
the $\beta$-functions for the Yukawa couplings $\beta_Y^{ijk}$
are given by
\be
\beta_Y^{ijk}
= Y^{p(ij}\ga^{k)}{}_p 
= Y^{ijp}\ga^k{}_p+(k\leftrightarrow i)+(k\leftrightarrow j),
\ee
where $\ga (g, Y)$ is the anomalous dimension for $\Phi$.
There exists an all-orders  relation
between the gauge $\beta$-function $\beta_g (g, Y)$ and  $\ga$ 
which was first derived
using  instanton calculus~\cite{nov}:
\be
\beta_g^{\NSVZ} =
{{g^3}\over{\lf}}\left[ {{Q- 2r^{-1}\Tr\left[\ga^{\NSVZ} C(R)\right]}  
\over{1- 2C(G)g^2{(\lf)}^{-1}}}\right].
\label{eq:russa}
\ee
Here $Q=T(R)-3C(G)$, 
$T(R)\delta_{ab} = \Tr(R_a R_b)$, $C(G)\delta_{ab} = f_{acd}f_{bcd}$, 
$r=\de_{aa}$ 
and $C(R)^i{}_j = (R_a R_a)^i{}_j$.

We have added a ``$\NSVZ$'' label to both $\beta_g$ and $\ga$ in 
Eq.~\ref{eq:russa} because of scheme dependence issues which we will discuss 
shortly. In the special case $Y=0$, the fixed point $g^*=0$, defined 
by 
\be
Q = \frac{2}{r}\Tr\left[\ga^{\NSVZ} C(R)\right]
\ee
is important for duality, in the context of the conformal window 
identified by Seiberg~\cite{seiberg}. We will return to this fixed point in the 
context of large-$N$ expansions. 

It turns out that if $\beta_g$ and $\ga$ are calculated using $\DRED$, 
then they begin to deviate from Eq.~\ref{eq:russa} at  three
loops~\cite{jjn}$^{\!,\,}$\cite{jjnb}. The relationship between 
$\beta_g^{\NSVZ}$ and $\beta_g^{\DRED}$ has been explored recently, 
with the conclusion that there exists an analytic redefinition  of $g$,
$g\to g'(g, Y)$ which connects them. We  emphasise that it
is quite non-trivial that the redefinition exists  at all; in the
abelian case for example, the redefinition consists  of a single term,
but it affects four distinct terms (with  different tensor structure) in
the $\beta$-functions.  By exploiting the  fact that  $N=2$ theories are
finite beyond one loop~\cite{hsw} it was  possible to  determine
$\beta_g^{\DRED}$ at three loops by a comparatively  simple calculation,
and at  four loops in the general case except for  one undetermined
parameter. What one learns from this is that it is highly  non-trivial
that the \DRED and \NSVZ results correspond to schemes which  can be
related in this manner. Use of what regularisation scheme 
would lead to the \NSVZ result, which is associated with the 
holomorphic nature of the Wilsonian action? 
Presumably, for example,  a combination 
of Pauli-Villars and higher derivatives~\cite{westb}.   
Notwithstanding the existence of the exact
\NSVZ result, however,   it is still important to have $\beta_g^{\DRED}$ as
accurately  as possible, because in calculating physical predictions 
\DRED  (or more accurately $\DREDD$) is the scheme most often used.  
The three loop results for $\beta_g^{\DRED}$ and $\ga^{\DRED}$ have found
phenomenological applications ~\cite{fjj}$^{\!,\,}$\cite{kolda}.  The
results for $\beta_g$ in \sic\ QCD (SQCD) 
with $N_f$ flavours and $N_c$ colours are:

\begin{equation}\begin{array}{rcl}
\lf\beta_g^{(1)} & = & \left(N_f - 3N_c\right)g^3, \\
& & \\
\llf\beta_g^{(2)} & = & \left(\left[4N_c-{2\over {N_c}}\right]N_f
-6N_c^2\right)g^5, \\
& & \\
\lllf\beta_g^{(3)} & = & \left(\left[{3\over{N_c}}-4N_c\right]N_f^2
+\left[21N_c^2-{2\over{N_c^2}}-9\right]N_f -21N_c^3
\right)g^7. \\
& & \end{array}
\label{eq:loopa}
\ee
  
For $\beta_g^{(4)}$ we have only a partial result~\cite{cjjc}:
\begin{equation}\begin{array}{rcl}
(16\pi^2)^4\beta_g^{(4)} & = &\biggl(-\frac{2}{3N_c}N_f^3
+\Bigl[\frac{100}{3}+4\alpha+\frac{6\kappa-20}{3N_c^2} 
- \bigl(\frac{62}{3}+2\kappa+8\alpha\bigr)N_c^2 \Bigr]N_f^2 \\
& + & \Bigl[36(1+\alpha)N_c^3-(34+12\alpha)N_c-\frac{8}{N_c}   
-\frac{4}{N_c^3}\Bigl]N_f \\ 
& - & (6+36\alpha)N_c^4  
\biggr)g^9 \\
& & \end{array} 
\label{eq:loopb}
\ee
where $\alpha$ is an as yet undetermined parameter, and where
$\kappa=6\zeta(3)$. (A recent application of the method of asymptotic 
Pad\'e approximants~\cite{mark} suggests that $\alpha\approx 2.4$.)

It is very interesting that the higher order group theory  invariants
found by  van Ritbergen et al~\cite{larin} in the corresponding
calculation for QCD do not appear here.  Of course the QCD calculation
was done with DREG rather than DRED;  but since these group structures
first appear at four loops we  would expect, for these particular
terms, that DRED and DREG should give the same  result at  this order. It
 is an excellent check on both calculations, therefore, that when in the
 QCD case we go to the special case of $N=1$ \sy\ (by setting  $N_f =
\frac{1}{2}$  and putting the fermions in the adjoint representation) 
the new invariants cancel. It is also interesting to note that in the 
pure gauge theory,   these invariants signalled the first contribution
from non-planar  structures to $\beta_g$ in QCD; for SQCD, it remains
possible  that $\beta_g^{\DRED}$ is free of such structures to all
orders  (this is manifestly so for $\beta_g^{\rm NSVZ}$ in the absence 
of chiral superfields, of course). 

Recently, some  exact results for soft supersymmetry-breaking masses and
couplings have been derived by Hisano and Shifman~\cite{shif} using the
holomorphy of  the Wilsonian action.  Soft breaking terms may be
accommodated within the superfield formalism by the  introduction of an
external ``spurion'' field~\cite{gir}, $\eta\equiv\theta^2$. The 
renormalisation-group functions for soft breaking parameters may all  be
derived  from the anomalous dimension $\gamma^{\eta}$ of the chiral
fields in the  presence of the spurion~\cite{yam}. $\gamma^{\eta}$ may
be expanded as 

\be 
\gamma^{\eta}=\gamma+\gamma^{[1]}\eta+\gamma^{[1]\dagger}\bar\eta
+\gamma^{[2]}\bar\eta\eta, \label{eq:spur} 
\ee where $\gamma$ is the
conventional anomalous dimension for $\Phi$, in the  absence of the
spurion. Furthermore, simple rules~\cite{yam} may be derived  for
obtaining $\gamma^{\eta}$ directly from $\gamma$. It  is possible to
derive from the exact results of Hisano and  Shifman~\cite{shif} an
elegant formula for the  $\beta$-function for the gaugino mass $M$,
namely 
\be \beta_M^{\NSVZ} = {2\over g}\left[ {M\beta^{\NSVZ}_g-
2g^3(\lf r)^{-1}\Tr\left[\ga^{[1]\NSVZ} C(R) \right]}\over{1-
2C(G)g^2{(\lf)}^{-1}}\right], \label{eq:spura} 
\ee where $\gamma^{[1]}$
is as defined in Eq.~\ref{eq:spur}. This result is strikingly similar in
form to the primordial NSVZ result for  $\beta_g$ in Eq.~\ref{eq:russa}.
 The other soft breaking $\beta$-functions are also simply  related to
$\ga^{[1]}$, as  follows (from now on we suppress the ``NSVZ'' label):
\be 
\beta_h^{ijk}= \ga^{(i}{}_l h^{jk)l} - 2\ga^{[1]}{}^{(i}{}_l
Y^{jk)l} \label{eq:spurb} 
\ee 
where $h^{ijk}\phi_i\phi_j\phi_k$ is the
soft $\phi^3$ interaction, and  
\be \beta_b^{ij} = \ga^{(i}{}_l b^{j)l}
-2\ga^{[1]}{}^{(i}{}_l \mu^{j)l} \label{eq:spurc} 
\ee 
where
$b^{ij}\phi_i\phi_j$ is the soft $\phi^2$ interaction. 

\section{Large-$N_f$ supersymmetric gauge theories}

The large-$N$ expansion is an alternative to conventional perturbation
theory. In both QCD and SQCD, the large $N_c$ expansion is
of particular interest~\cite{witt}; more tractable, however, 
is the large $N_f$ expansion. Recently the 
leading and $O(1/N_f)$ terms in $\beta_g$ and $\ga$ have been calculated 
(using DRED) 
for a number  of \sic\ theories~\cite{fjjn}. We give below the results for 
SQCD (noting that we have rescaled the gauge coupling, $g\to g/\sqrt{N_f}$):
\be
\ga
= -\frac{(N_c^2 - 1)}{N_fN_c}\Khat G(\Khat), 
\ee
and
\begin{equation}\begin{array}{rcl}
\beta_g & = &g\Khat -\frac{3N_c}{N_f}g\Khat
+4g\Khat\frac{N_c}{N_f}\int_{0}^{\hat K}
(1-x)G(x)\,dx
\\
& & \\
\qquad & - & 
\frac{2g\Khat}{N_c N_f}\int_{0}^{\hat K}(1-2x)G(x)\,dx . \\
& & \end{array}
\label{eq:bign}
\ee
where $\Khat = g^2/(16\pi^2)$, and 
\be
G(x) = \frac{\Gamma (2 -2x)}{\Gamma(2-x)
\Gamma(1-x)^2\Gamma(1+x)}.
\ee  
These results do not satisfy Eq.~\ref{eq:russa}, because they 
were calculated using DRED. 
It is quite remarkable that the $O(1/N_f)$ corrections
to the SQCD $\beta$-function depend only on simple integrals
involving $G(x)$. $G$ has a simple pole at $x=3/2$ and
consequently $\beta_g$ has a logarithmic singularity at
$g^2 =24\pi^2$ and a finite radius of convergence in $g$. 
Using Eq.~\ref{eq:bign} for $N_f=6$ and  $N_c =3$, values which lie in 
the conformal window $3N_c/2 < N_f <3N_c$, we indeed find 
an infra-red fixed point in the gauge coupling evolution, corresponding 
to $g^*\approx 8$. It is interesting that the range of $N_f$ such 
that $g^* < 24\pi^2$ is given by $$aN_c < N_f < 3N_c,$$ where $a\approx 1.7$ 
depends weakly on $N_c$. This is remarkably close 
to the exact conformal window. 

Of course the result for $g^*$ is scheme dependent. 
In the NSVZ scheme it is possible to show that the $O(1/N_f)$ contribution 
to $\ga$ is in fact the same as in DRED, with the corresponding result 
for $\beta_g$ being easily calculated from Eq.~\ref{eq:russa}. One 
then finds that the fixed point (for $N_c=3$) corresponds to $g^*\approx 7$.

Of course it is not clear that  this regime is within the region of
validity of our approximation: do we believe that the  appropriate
expansion parameter is $N_c/N_f$ or $3N_c/N_f$?  It would obviously be
interesting if we could  calculate more terms in the $1/N_f$ expansion. 
Even the $O(1/N_f^2)$ contribution presents considerable technical
problems, however. 





\end{document}